\documentclass[review,12pt]{elsarticle}

\usepackage{amssymb}
\usepackage{amsthm}
\usepackage{amsmath}
\usepackage{mathrsfs}
\usepackage{graphicx}
\usepackage{epstopdf}
\usepackage{float}
\usepackage{caption}
\usepackage{subcaption}
\usepackage{booktabs}
\usepackage{natbib}
\usepackage{bm}
\usepackage{mathrsfs}
\usepackage{soul}
\usepackage{graphicx}
\usepackage{xcolor}
\usepackage{courier} 
\usepackage{listings} 
\usepackage{tabu} 
\usepackage{longtable}
\usepackage{changepage} 
\usepackage[margin=2.2cm]{geometry}
\biboptions{sort&compress} 
\usepackage{hyperref}
\usepackage{cleveref}

\journal{Materials \& Design}

\makeatletter
\def\@author#1{\g@addto@macro\elsauthors{\normalsize%
    \def\baselinestretch{1}%
    \upshape\authorsep#1\unskip\textsuperscript{%
      \ifx\@fnmark\@empty\else\unskip\sep\@fnmark\let\sep=,\fi
      \ifx\@corref\@empty\else\unskip\sep\@corref\let\sep=,\fi
      }%
    \def\authorsep{\unskip,\space}%
    \global\let\@fnmark\@empty
    \global\let\@corref\@empty  
    \global\let\sep\@empty}%
    \@eadauthor={#1}
}
\makeatother

\Crefname{equation}{Eq.}{Eqs.}
\Crefname{figure}{Fig.}{Figs.}

\renewcommand{\baselinestretch}{1.} 

\begin{document}

\begin{frontmatter}

\title{A computational framework to predict weld integrity and microstructural heterogeneity: application to hydrogen transmission}

\author[OXFORD]{Job Wijnen}

\author[EPRI]{Jonathan Parker}

\author[EPRI]{Michael Gagliano}

\author[OXFORD]{Emilio Mart\'{\i}nez-Pa\~neda\corref{cor1}}
\ead{emilio.martinez-paneda@eng.ox.ac.uk}

\address[OXFORD]{Department of Engineering Science, University of Oxford, Oxford OX1 3PJ, UK}
\address[EPRI]{Electric Power Research Institute, 3420 Hillview Avenue, Palo Alto, CA 94304, USA}

\cortext[cor1]{Corresponding author.}

\begin{abstract}

\noindent We present a novel computational framework to assess the structural integrity of welds. In the first stage of the simulation framework, local fractions of microstructural constituents within weld regions are predicted based on steel composition and welding parameters. The resulting phase fraction maps are used to define heterogeneous properties that are subsequently employed in structural integrity assessments using an elastoplastic phase field fracture model. The framework is particularised to predicting failure in hydrogen pipelines, demonstrating its potential to assess the feasibility of repurposing existing pipeline infrastructure to transport hydrogen. First, the process model is validated against experimental microhardness maps for vintage and modern pipeline welds. Additionally, the influence of welding conditions on hardness and residual stresses is investigated, demonstrating that variations in heat input, filler material composition, and weld bead order can significantly affect the properties within the weld region. Coupled hydrogen diffusion-fracture simulations are then conducted to determine the critical pressure at which hydrogen transport pipelines will fail. To this end, the model is enriched with a microstructure-sensitive description of hydrogen transport and hydrogen-dependent fracture resistance. The analysis of an X52 pipeline reveals that even 2 mm defects in a hard heat-affected zone can drastically reduce the critical failure pressure. 

\end{abstract}


\end{frontmatter}


\section{Introduction}
\label{section:introdcution}

Understanding the interplay between process parameters, material heterogeneity and fracture resistance is key to assessing the structural integrity of components resulting from welding, additive manufacturing and other fabrication processes \cite{Dong2024,Zinovieva2024,Wagner2024,Sowards2015}. Moreover, a mechanistic understanding of such interplay would enable an optimized approach for these and similar processes. This necessitates establishing a new class of thermo-metallurgical-mechanical models that can integrate coupled process and fracture modeling.

One area where mechanistic, integrated process-fracture modeling can be particularly valuable is pipeline integrity assessment, with applications across energy distribution, where welds constitute the weakest link. Existing gas transmission pipeline networks are very diverse, with pipelines manufactured from different steel grades and installed over several decades. Most pipeline steels are categorized according to the API 5L specification, with in-service pipelines mainly ranging from X42 to X80 grades, where the number denotes the minimum yield strength in ksi units. These different steel grades may exhibit distinct microstructures, which can influence in-service performance and in particular the fracture resistance. The vast majority of pipeline failures do not occur in the base material but in the weld region of the pipeline. The weld metal (WM) and surrounding heat-affected zone (HAZ) may be subjected to multiple cycles of rapid heating and subsequent cooling, resulting in residual stresses and the possible formation of harder phases, such as bainite or martensite. Consequently, these regions are susceptible locations of a pipeline network and defects in these locations, introduced during production, welding, or during service time, can lead to catastrophic failures \cite{Vishnuvardhan2023,Goodfellow2023}. Given the variability in pipeline materials, welding techniques, and size and distribution of defects, assessing the performance of existing pipeline infrastructure is a major challenge, underscoring the potential of numerical approaches that can quickly and cheaply evaluate the risk of different scenarios. This is now urgently needed to assess whether the existing gas pipeline infrastructure can be repurposed to transport hydrogen gas, to support zero carbon energy commitments \cite{Laureys2022}. Hydrogen is known to significantly degrade metals and alloys through a phenomenon termed \emph{hydrogen embrittlement} \cite{Djukic2019,chen2024hydrogen}, and it has been experimentally demonstrated that pipeline steels can undergo very significant reductions in ductility, fracture toughness and fatigue crack growth resistance when exposed to hydrogen-containing environments \cite{Ronevich2021,Bortot2024,Jemblie2024}. Given computational methodological developments in the residual stress and structural integrity communities, there is an opportunity to utilize these to better characterize the behavior of hydrogen transmission pipelines. 

Numerical welding models, often combined with experimental data, have been used to characterize residual stress patterns for different welding processes \cite{Song2013,Dong2007,Pissanti2019,Bouchard2007,Song2015}. To accurately predict residual stresses in ferritic steel welds, several models have included solid-state phase transformations to predict the presence of hard phases in the weld region \cite{Jiang2018,Bok2011,Hamelin2014,Sun2019,Sun2019b}. However, to develop a complete structural integrity framework for the accurate assessment of pipelines, welding process predictions need to be integrated with failure process simulations. This is not straightforward, as it requires integrating process modeling with a computational fracture method capable of capturing crack nucleation and growth along non-predefined paths, and incorporating local material heterogeneities and residual stresses in a mesh-independent setting.

The recent development of phase field fracture methods provide a suitable pathway to achieve this objective. Phase field approaches to fracture have been shown to be capable of capturing fracture problems of arbitrary complexity, including crack coalescence and branching, in 2D and 3D, for both mechanical and coupled problems, and in homogeneous and non-homogeneous solids \cite{natarajan2019phase,bourdin2014morphogenesis,boyce2022cracking}. Very recently, Mandal \textit{et al.}\ \cite{Mandal2024} combined welding process modeling with a phase field-based coupled deformation-diffusion-fracture model, and applied this framework to predict the critical pressure at which seam welds of hydrogen transport pipelines would fail. Their results revealed that microstructural heterogeneities and residual stresses can have a very significant impact on the structural integrity of hydrogen transport pipelines. However, the welding process model used did not resolve phase changes and therefore microstructural heterogeneities had to be introduced through image processing of experimental microhardness maps. This is a notable drawback as the ability of models to handle a wide range of scenarios is now limited by the availability of microhardness maps. In addition, determining local fracture properties based on variations in microhardness comes with considerable uncertainty.

\begin{figure}[!tb]
	\centering 
	\includegraphics[width=\linewidth]{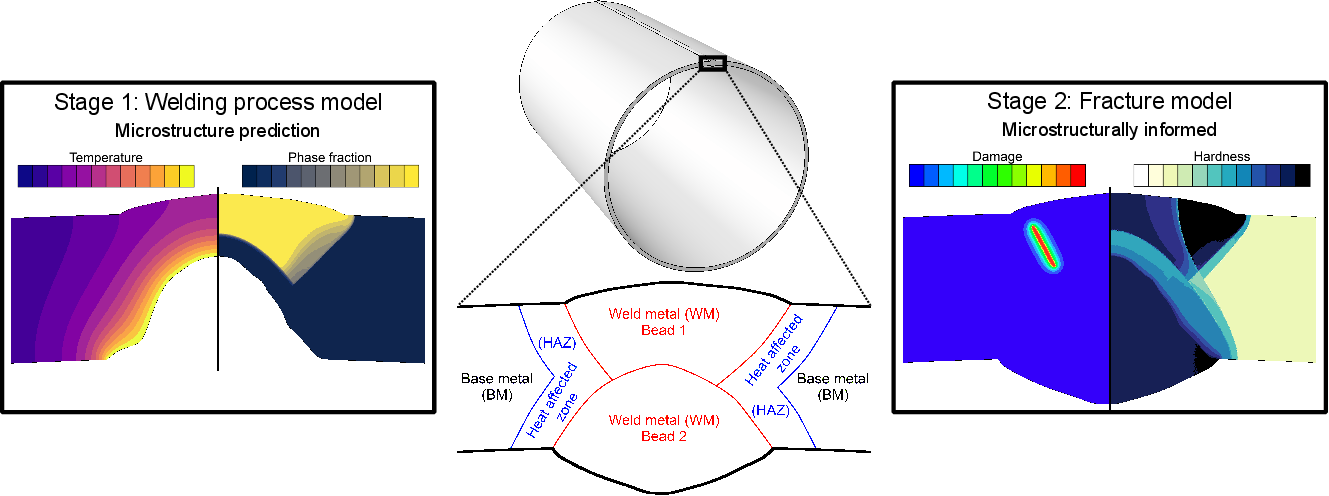}
	\caption{Schematic of the different zones in a weld and outline of the two-stage modeling framework. In stage 1, the welding process, including solid-state phase transformations in the underlying microstructure, is simulated. In stage 2, welding process results are used as initial conditions for an elastoplastic phase field fracture model.}
	\label{fig:frameworkoutline}
\end{figure}

The objective of this paper is to develop a comprehensive structural integrity framework that is able to predict microstructural heterogeneity and potential susceptible locations within the weld and subsequently predict pipeline failure. The framework, outlined in \Cref{fig:frameworkoutline}, combines for the first time microstructurally-sensitive process modeling with phase field fracture simulations. The first stage involves incorporating predictive capabilities of local microstructural constituents based on steel composition, weld process details, and associated thermal cycles using Li's phenomenological model \cite{Li1998}. Furthermore, phase transformations lead to additional strains, which are incorporated into the residual stress calculation. Critical outputs of this first step are local phase fraction maps within the weld metal and heat affected zones. In the second stage of the framework, the phase fraction maps will be used to determine the heterogeneity in material properties, which are then used for structural integrity simulations using an elastoplastic phase field fracture model. To assess the effect of hydrogen on the structural integrity of pipelines, a hydrogen diffusion model is coupled to the phase field fracture model, where the  hydrogen concentration locally reduces the fracture toughness of the material. The resulting weld process-fracture model can therefore predict the structural integrity of hydrogen transport pipelines for any choice of weld configuration and protocol, defect distribution and hydrogen partial pressure.

The remainder of this paper is structured as follows. In \Cref{section:model}, the simulation framework is described, spanning both the thermo-metallurgical-mechanical weld process model and the subsequent coupled phase field-based deformation-diffusion-fracture stage. Next, \Cref{section:validation} demonstrates the enhanced welding process model and validates its predictions against experimentally determined microhardness maps. The potential of the framework in predicting the failure of welded components is addressed in \Cref{section:application}, where the behavior of an X52 hydrogen transport pipeline is evaluated under different scenarios. Finally, the manuscript ends with a summary and concluding remarks in \Cref{Sec:Conclusions}. 

\section{Modeling framework}
\label{section:model}

This section describes the modeling framework, which consists of two stages, as illustrated in \Cref{fig:frameworkoutline}. The first stage of the framework, described in \Cref{section:weldingmodel}, involves the use of a thermal-metallurgical-mechanical model to simulate the welding process. First, the metallurgical model, governing solid-state phase transformations (SSPT), is outlined. Next, the integration into the heat transfer finite element model is described. Finally, the calculation of associated residual stresses is explained. In stage 2 of the simulation framework, an elastoplastic phase field model is used to simulate deformation and failure of the pipeline subjected to internal (hydrogen) pressure, as described in \Cref{section:fracturemodel}. The model uses results from the first stage of the framework as initial conditions.

\subsection{Welding process model}
\label{section:weldingmodel}

A weld geometry can commonly be divided into three regions, as schematically shown in \Cref{fig:frameworkoutline}. The base metal (BM) refers to the original pipeline material unaffected by thermal cycles associated with the welding process.  In the submerged arc welding process considered in this study, weld beads, consisting of a filler material, are deposited to join the base metal on both sides of the weld. This region with filler material is referred to as weld metal (WM) and is typically characterized by a microstructure with large columnar grains that grow from the fusion line to the centerline of the weld bead, driven by the high temperature gradient. The heat affected zone (HAZ) refers to the small regions adjacent to the WM, where the original pipeline material is heated to such an extent that microstructural changes, such as phase transformation or recrystallization, occur. For more details on various welding processes and the resulting microstructures, the reader is referred to relevant literature \cite{Lancaster1999}.

The model described in this section aims to predict the phase transformations and accompanying residual stresses in the WM and HAZ. The model predicts the microstructure at the macroscopic scale.  This means that each continuum material point is assumed to have an underlying microstructure that can consist of multiple phases, which are described in terms of phase fractions. These phase fractions directly affect local material properties and stress responses. Other microstructural features, such as grain morphology, are not taken into account. 

\subsubsection{Metallurgical model}
\label{section:transformationmodel}

The solid-state phase transformations of steel involve many complex processes. On the level of the microstructure, multi-phase-field methods have proven to be a powerful tool for simulating microstructure evolution during phase transformations of steels \cite{Steinbach2009,Militzer2011,Yamanaka2023,Liang2023}. However, such simulations are not computationally tractable on the scale of full welds. Therefore, semi-empirical models are commonly adopted to describe the transformation kinetics at the macroscale. 

When carbon steels are heated to a sufficiently high temperature, their microstructure transforms into austenite. Upon subsequent cooling, austenite can decompose into ferrite, pearlite, bainite, or martensite, depending on the cooling rate and alloying element content \cite{Bhadeshia2006}. This decomposition can be categorized into two types. The first three transformations are considered diffusive, while the last one is displacive. Diffusive transformations are characterized by sufficiently slow cooling rates, allowing atoms to diffuse and re-order. In displacive transformations, the cooling rate is so fast that atoms do not have time to re-order. In this case, the transformation is accompanied by a shear deformation of the crystal lattice.

The slow cooling during the fabrication of steel pipelines typically results in a mixed microstructure of ferrite and pearlite \cite{Agnani2023,Zhu2015}. During the welding process, the high temperatures in and close to the WM result in a fully austenitic microstructure in these regions. Due to the relatively fast cooling rates after the welding process, this austenite can transform into harder phases like bainite or martensite \cite{Huda2021,Zhu2015,Lancaster1999}. In the HAZ, peak temperatures during the welding process are lower than in the WM, resulting in only partial re-austenitization of the original microstructure. Consequently, the HAZ commonly consists of a mixture of softer phases and harder phases. Additionally, heat accumulation resulting from multiple thermal cycles can lead to more heterogeneous and defective HAZ microstructures \cite{Gao2022,Guirguis2024}.

The models describing the diffusive decomposition of austenite into ferrite, pearlite, or bainite can be broadly subdivided into two categories, namely, Kirkaldy type models \cite{Kirkaldy1983,Li1998} and Johnson-Mehl-Avrami-Kolmogorov type (JAMK) models \cite{Johnson1939,Avrami1939,Fanfoni1998,Blazquez2022}. JMAK models require extensive fitting of experimental data \cite{Bok2015,Ahn2017,Li2016}, whereas empirical relations are available for most parameters of Kirkaldy models. Here, we adopt the Kirkaldy-type model of \cite{Li1998}. The suitability of Li's model to predict phase transformations and accompanying residual stresses during welding simulations or continuous cooling transformation (CCT) diagrams has been validated by several studies \cite{Bok2011,Hamelin2014,Sun2019,Sun2019b,Collins2023}.

The general form of the Kirkaldy model for isothermal transformations is given by
\begin{equation}
	\tau(X, T) = \frac{F(\text{C,Mn,Si,Ni,Cr,Mo,G})}{\Delta T^n \exp(-Q/(RT))}S(X).
	\label{eq:transf_general}
\end{equation}
Here, $\tau$ is the time it takes for a transformation to reach a phase fraction $X$ at temperature $T$, $\Delta T$ is the undercooling temperature, $Q$ the activation energy, $R$ the universal gas constant, $F$ is a function of the chemical composition and the ASTM prior austenite grain size number $G$, and $S$ is a sigmoidal function. Li \textit{et al.}\ \cite{Li1998} fitted this equation to data from the literature for ferrite, pearlite, and bainite, resulting in
\begin{equation}
	\tau_f(X,T) = \frac{F_f}{2^{0.41G} (A_{e3}-T)^3\exp(-13840/T)}S(X),
	\label{eq:transf_f}
\end{equation}
\begin{equation}
	\tau_p(X,T) = \frac{F_p}{2^{0.32G} (A_{e1}-T)^3\exp(-13840/T)}S(X),
	\label{eq:transf_p}
\end{equation}
and
\begin{equation}
	\tau_b(X,T) = \frac{F_b}{2^{0.29G}(B_{s}-T)^2\exp(-13840/T)}S(X),
	\label{eq:transf_b}
\end{equation}
with $A_{e3}$, $A_{e1}$, and $B_s$ being, respectively, the ferrite, pearlite, and bainite transformation start temperatures. The compositional constants are given by
\begin{equation}
	F_f = \exp(1 + 6.31\text{C} + 1.78\text{Mn} + 0.31\text{Si} + 1.12\text{Ni} + 2.70\text{Cr} + 4.06\text{Mo}),
        \label{eq:compfunc_f}
\end{equation}
\begin{equation}
	F_p = \exp(-4.25 + 4.12\text{C} +4.36\text{Mn} + 0.44\text{Si} + 1.71\text{Ni} + 3.33\text{Cr} + 0.36\text{Mo}),
        \label{eq:compfunc_p}
\end{equation}
\begin{equation}
	F_b = \exp(-10.23 + 10.18\text{C} + 0.85\text{Mn} + 0.55\text{Ni} + 0.90\text{Cr} + 0.36\text{Mo}),
        \label{eq:compfunc_b}
\end{equation}
where the alloying elements are defined in wt.\%.
Finally, the sigmoidal function describing the behavior of the transformation over time is defined as
\begin{equation}
	S(X) = \int_0^{X} \frac{1}{X^{0.4(1-X)}(1-X)^{0.4X}} \, dX.
    \label{eq:sigmoid}
\end{equation}

\begin{figure}[!tb]
    \begin{subfigure}{\linewidth}
        \centering
        \includegraphics[width=\linewidth]{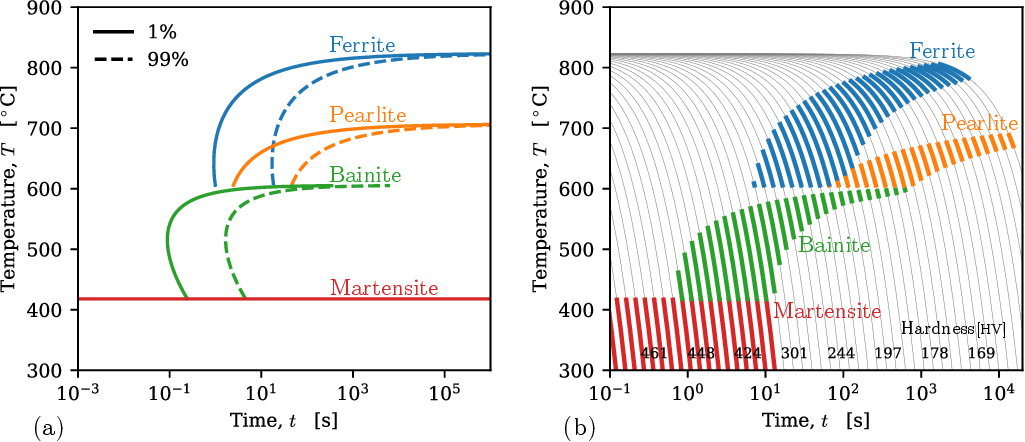}
        \phantomcaption \label{fig:x60_ttt}
        \phantomcaption \label{fig:x60_cct}
    \end{subfigure}
    \caption{Phase  transformation behavior of the implemented model. (a) Time-temperature-transformation diagram obtained by \Cref{eq:transf_f,eq:transf_p,eq:transf_b}. (b) Continuous cooling transformation diagram obtained by making use of the additivity rule.}
    \label{fig:x60_ttt_cct}
\end{figure}

\Cref{eq:transf_f,eq:transf_p,eq:transf_b} give the isothermal transformation times within the temperature range for which a particular phase transformation is active, and can directly be used to plot the time-temperature transformation (TTT) diagram, as shown in \Cref{fig:x60_ttt}. Here the solid lines denote the times at which 0.1\% of that phase has formed, while the dashed lines represent the times at which 99.9\% of the phase has formed. To obtain phase fractions for a non-isothermal temperature path, the additivity rule can be used \cite{Scheil1935}. A final phase fraction can then be obtained by
\begin{equation}
    X = \int_0^{\tau^*} \frac{dX}{dt} dt.
    \label{eq:additivity}
\end{equation}
The general form of the phase fraction rates, derived from \Cref{eq:transf_general}, is
\begin{equation}
	\frac{dX}{dt} = \frac{\Delta T^n \exp(-Q/(RT))}{F} X^{0.4(1-X)}(1-X)^{0.4X}.
	\label{eq:diff}
\end{equation}

It should be noted that when the initial phase fraction is 0, no transformation takes place according to \Cref{eq:diff}. Therefore, following Hamelin \textit{et al.}\ \cite{Hamelin2014}, a nucleation phase is introduced. When a phase has a fraction of 0, it will be nucleated at time $t_n$ with a fraction of 0.01 when the following equation is satisfied:
\begin{equation}
	\int_0^{t_n} \frac{1}{\tau(0.01,T)} \, dt = 1.
	\label{eq:nucleation}
\end{equation}

With \Cref{eq:additivity,eq:nucleation}, any cooling path can be simulated. By calculating the austenite decomposition for different, but constant cooling rates, continuous cooling transformation (CCT) diagrams can be constructed. An example is depicted in \Cref{fig:x60_cct}, where the temperature paths of different cooling rates are shown. For each path, the colored parts indicate the times at which a certain transformation is active.

The maximum attainable ferrite and pearlite fractions are limited by thermodynamic constraints. Following Bok \textit{et al.}\ \cite{Bok2011}, the constrained ferrite and pearlite fractions are calculated through
\begin{equation}
	X_f^\text{true} = X_f X_f^\text{eq}
\end{equation}
and
\begin{equation}
	X_p^\text{true} = X_p \left( 1 - X_f^\text{eq} \right),
\end{equation}
where $X_f^\text{eq}$ is the equilibrium ferrite phase fraction.

In carbon steels, the martensite transformation is instantaneous, i.e.\ it is time-independent and only depends on the temperature. This transformation is commonly described by the Koistenen-Marburger equation \cite{Koistinen1959}. However, since no martensite formation is observed in any of the results presented in this paper, i.e.\ all the austenite has decomposed into other phases before the martensite transformation temperature is reached, the details will not be described here.

Since the material in the HAZ goes through multiple thermal cycles, the formation of austenite upon heating also needs to be described by the model. The kinetic equation introduced by Leblond and Devaux \cite{Leblond1984} is employed here. It is given as
\begin{equation}
	\frac{dX_a}{dt} = \frac{X_a^\text{eq} - X_a}{\tau_{LD}},
\end{equation}
where $X_a^\text{eq}$ is the equilibrium fraction and $\tau_{LD}$ is the characteristic transformation time. Following Sun \textit{et al.}\ \cite{Sun2019}, $X_a^\text{eq}$ increases linearly from 0 to 1 between temperatures $A_{e1}$ and $A_{e3}$, while $\tau_{LD}$ decreases linearly from 1 to 0.05.

The diffusive phase transformations of \Cref{eq:transf_f,eq:transf_p,eq:transf_b} are dependent on the prior austenite grain (PAG) size number $G$. Since this PAG size can vary significantly throughout the HAZ and the weld zone, austenite grain growth is accounted for in the model. Here, the model by Pous-Romero \textit{et al.}\ \cite{PousRomero2013} is adopted. The evolution of the average austenite grain diameter, $D$, is described as
\begin{equation}
	\frac{dD}{dt} = A \exp \left( \frac{-Q}{RT} \right) \left( \frac{1}{D} - \frac{1}{D_\text{lim}} \right),
	\label{eq:graingrowth}
\end{equation}
where $Q$ is the activation energy,  $A$ is a material parameter and $D_\text{lim}$ is the maximum grain diameter. This limit on grain diameter represents the effect of precipitation, which limits grain growth. An activation energy of $Q=190$  kJ is adopted \cite{PousRomero2013}.

When the austenite formation starts, the austenite grain size is set to an initial grain size $D_0$, after which it evolves during the heating stage according to \Cref{eq:graingrowth}. The conversion from grain diameter, $D$, to ASTM grain size number, $G$, is given by \cite{ASTME112}
\begin{equation}
	G = 2 \log_2 \left(254/D\right) + 1.0.
\end{equation}

\subsubsection{Thermal model}
\label{section:thermalmodel}

The heat transfer during the welding process is described by
\begin{equation}
	c(T,X_i) \rho(T,X_i) \frac{dT}{dt} = - \nabla \cdot \mathbf{q},
\end{equation}
with specific heat $c$, density $\rho$ and heat flux
\begin{equation}
	\mathbf{q} = -k(T,X_i) \nabla T,
\end{equation}
with thermal conductivity $k$. The material properties $c$, $\rho$, and $k$ are both temperature and phase-dependent. A linear rule-of-mixtures is used to determine the properties of a multi-phase microstructure. For example, the conductivity is calculated by
\begin{equation}
	c(T,X_i) = \sum_{i=f,p,b,m,a} X_i c_i(T).
	\label{eq:ruleofmixtures}
\end{equation}

Convective and radiative boundary conditions are used to simulate the heat loss to the surroundings:
\begin{equation}
	\mathbf{q} \cdot \mathbf{n} = q_c + q_h \quad \text{on } \Gamma_q,
\end{equation}
where $\mathbf{n}$ is the surface normal. Here, $q_c$ is the convective heat flux is given by
\begin{equation}
	q_c = h_c \left( T - T_0 \right),
\end{equation}
with $h_c$ being the convective heat transfer coefficient and $T_0$ being the ambient temperature. The radiative heat flux is given by
\begin{equation}
	q_r = \epsilon\sigma \left( (T - T_{Z})^4 - (T_0 - T_{Z})^4 \right),
\end{equation}
where $\epsilon$ is the emissivity of the surface, $\sigma$ is the Stefan-Boltzmann constant, and $T_Z$ is the absolute zero temperature.

\subsubsection{Residual stress calculation}
\label{section:mechanicalmodel}

Constrained thermal expansions and contractions, along with phase transformation strains occurring during the welding process lead to residual stresses in the pipeline. To model the mechanical deformation, a small strain formulation is employed in which the strain of the material is additively decomposed into thermal ($th$), phase transformation ($tr$), elastic ($e$), plastic ($p$), and transformation-induced plasticity ($tp$) strains:
\begin{equation}
	\bm{\varepsilon} = \bm{\varepsilon}^{th} + \bm{\varepsilon}^{tr} + \bm{\varepsilon}^{e} + \bm{\varepsilon}^{p} + \bm{\varepsilon}^{tp}.
\end{equation}

Elastic deformations are related to the stress state via linear isotropic elasticity:
\begin{equation}
	\bm{\sigma} = \kappa \varepsilon^{e}_\text{vol} \bm{I} + 2G\bm{\varepsilon}^{e}_\text{dev},
\end{equation}
where $\varepsilon^\square_\text{vol}=\text{trace}(\bm{\varepsilon}^\square)$ and $\bm{\varepsilon}^\square_\text{dev}=\bm{\varepsilon}^\square - \tfrac{1}{3}\varepsilon^{\square}_\text{vol}\bm{I}$ denote, respectively, the volumetric and deviatoric components of a strain tensor, $\kappa$ is the bulk modulus, and $G$ is the shear modulus. The stress state must satisfy Cauchy's momentum balance:
\begin{equation}
	\nabla \cdot \bm{\sigma} = \mathbf{0}.
\end{equation} 

\begin{figure}[!tb]
	\begin{subfigure}{\linewidth}
		\centering 
		\includegraphics[width=\linewidth]{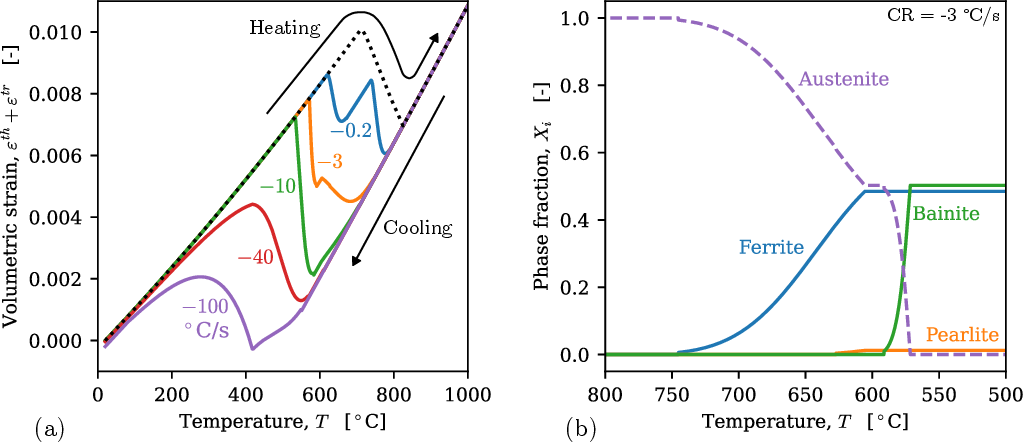}
		\phantomcaption \label{fig:expansion_strain}
		\phantomcaption \label{fig:expansion_phases}
	\end{subfigure}
	\caption{Phase transformation behavior upon cooling down from the fully austenitic phase. (a) Volumetric strain due to thermal expansion and phase transformations for thermal cycles with different cooling rates. (b) Evolution of the phase fractions upon cooling with a cooling rate of -3$^\circ$C/s.}
	\label{fig:expansion}
\end{figure}

Besides conventional thermal expansion, phase transformations lead to additional volumetric expansion or shrinkage. Following the model by Jablonka \textit{et al.}\ \cite{Jablonka1991}, the combined thermal and phase transformation strains are calculated based on the density of the material:
\begin{equation}
	\bm{\varepsilon}^{th} + \bm{\varepsilon}^{tr} = \left( \sqrt[3]{ \frac{\rho_\text{ref}}{\rho(T,X_i)} } -1 \right) \bm{I}.
\end{equation}
Here, $\rho(T,X_i)$ is the density of the steel, calculated via the linear rule of mixtures (\Cref{eq:ruleofmixtures}), and $\rho_\text{ref}$ is the initial reference density. \Cref{fig:expansion_strain} shows the total volumetric expansion during thermal cycles with different cooling rates. For each rate, phase transformations take place at varying temperatures. This leads to different volumetric expansion curves. For example, the evolution of the phase fractions during the cooling stage with a rate of -3$^\circ$C/s are shown in \Cref{fig:expansion_phases}.

In addition to volumetric strains, phase transformations can introduce additional deviatoric plastic strains. These so-called transformation-induced plastic (TRIP) strains can occur at stress levels lower than the macroscopic yield stress. Two mechanisms are commonly identified, namely, the Greenwood and Johnson mechanism \cite{Greenwood1965} and the Magee and Paxton mechanism \cite{Magee1966}. In the former, the dilation of the material due to phase transformations causes additional stresses that can cause plastic deformation. In the latter, an applied stress promotes martensite to form in a preferred crystallographic direction, resulting in a non-zero deviatoric eigenstrain. On a macroscopic scale, the Greenwood and Johnson mechanism is usually considered dominant and is accounted for through the empirical relation \cite{Fischer1996}:
\begin{equation}
	\dot{\bm{\varepsilon}}^{tp} = \sum_{i=f,p,b,m} 3 K_{a \rightarrow i} \left( 1 - X_i \right) \left< \dot{X}_i \right> \bm{\sigma}_\text{dev},  
\end{equation}
where $\bm{\sigma}_\text{dev}=\bm{\sigma} - \tfrac{1}{3}\text{trace}(\bm{\sigma})\mathbf{I}$ is the deviatoric part of the stress tensor, $\left< \square \right>$ denote Macaulay brackets, and $K_{a\rightarrow i}$ are the TRIP parameters.

Isotropic von Mises plasticity with power law hardening is used. The plastic flow relation is given by
\begin{equation}
	\dot{\bm{\varepsilon}}^p = \dot{\varepsilon}^p_\text{eq}  \frac{\partial f}{\partial \bm{\sigma}},
	\label{eq:plasticflow}
\end{equation}
with yield surface
\begin{equation}
	f = \sigma_\text{eq} - \sigma_{y0}\left(1 + \frac{E (\varepsilon^p_\text{eq}+\varepsilon^{tp}_\text{eq})}{\sigma_{y0}}\right)^n,
	\label{eq:yieldsurface}
\end{equation}
subjected to the Kuhn-Tucker constraint conditions
\begin{equation}
	\dot{\varepsilon}^p_\text{eq} \geq 0, \quad f \leq 0, \quad \dot{\varepsilon}^p_\text{eq}f=0,
	\label{eq:plasticconstraints}
\end{equation}
where $\varepsilon^\square_\text{eq} = \sqrt{\tfrac{2}{3} \bm{\varepsilon}^\square_\text{dev} : \bm{\varepsilon}^\square_\text{dev}}$ denotes an equivalent strain, $\sigma_\text{eq} = \sqrt{\tfrac{3}{2} \bm{\sigma}_\text{dev} : \bm{\sigma}_\text{dev}}$ is the equivalent von Mises stress, $\sigma_{y0}$ is the initial yield strength, and $n$ is the hardening exponent. Note that the hardening of a material is governed by the total plastic deformation, which includes the TRIP strain.

When metals and alloys reach high temperatures the thermal energy can reduce welding residual stress values. To capture this annealing effect, the accumulated plastic strain, $\varepsilon^p_\text{eq}$, is reset to 0 for high temperatures \cite{Smith2012,Muransky2015}. This eliminates the strain hardening history of the material. Following the recommendation of Muránsky \textit{et al.}\ \cite{Muransky2015} for isotropic plasticity,  a single annealing temperature, $T^a$, that is close to the melting temperature of the steel is adopted.

\subsubsection{Weld model parameters}

Key parameters of the transformation models described in Section \ref{section:transformationmodel} are the transformation temperatures $A_{e1}$, $A_{e3}$, $B_s$, and $M_s$. These can be estimated using empirical formulae based on steel composition, as reported in literature \cite{Andrews1965,Grange1961}. Alternatively, thermodynamic software can be used to obtain these temperatures \cite{Saunders2003}. In the current study, the latter approach is used. Thermodynamic software can also be employed to obtain temperature-dependent properties of individual phases. For the heat transfer part of the simulation, the required properties are the density $\rho$, thermal conductivity $k$, and specific heat $c$. As previously explained, properties in multi-phase regions are computed by averaging the individual phases, in accordance with \Cref{eq:ruleofmixtures}. In terms of the mechanical part of the simulation, relevant properties are the phase and temperature-dependent bulk moduli ($\kappa$), shear moduli ($G$), initial yield stresses ($\sigma_{y0}$), and TRIP parameters ($K_{a \rightarrow i}$).

\subsection{Fracture model for hydrogen embrittlement}
\label{section:fracturemodel}

The coupled deformation-diffusion-fracture phase field model for elastic-plastic solids exposed to hydrogen-containing environments is presented here. 

\subsubsection{Elastoplastic phase field fracture}

Phase field fracture modeling is used to describe the evolution of cracks, as it has shown to be physically sound, based on the thermodynamics of fracture \cite{bourdin2000numerical,kristensen2021assessment}, proven to be capable of blindly predicting experiments \cite{navidtehrani2024damage,wu2021three}, computationally robust, exhibiting no convergence issues \cite{kristensen2020phase,khimin2022space} and capturing very complex cracking phenomena \cite{Borden2016,bourdin2014morphogenesis}. Moreover, of critical importance here, the material toughness $G_c$ is defined at the integration point level and therefore can be made phase- and hydrogen concentration-dependent. 

The present phase field fracture implementation follows largely the variationally-consistent elastic-plastic phase field fracture model presented in Refs. \cite{Mandal2024,cupertino2024suitability}. In the phase field fracture paradigm, an order parameter, $\phi$, is used to regularize discrete cracks. A value of $\phi=1$ represents fully damaged material, while a value of $\phi=0$ represents fully intact material. Upon adopting a suitable energy density functional, a smooth transition between intact and damaged regions is obtained. Here, the following energy functional of the material, defined over domain $\Omega$, is used
\begin{equation}
	\Psi = \int_\Omega \psi + \gamma(\phi) G_c(C_L) \, d\Omega,
	\label{eq:functional}
\end{equation}
in which $\psi$ is the strain energy density, $G_c$ is the critical energy release rate which depends on the local hydrogen concentration $C_L$, and $\gamma$ is the crack regularization function given by \cite{bourdin2000numerical}
\begin{equation}
	\gamma = \frac{1}{2\ell}\left(\phi^2 + \ell^2 |\nabla \phi|^2 \right),
	\label{eq:crackfunc}
\end{equation}
with $\ell$ being a length scale parameter governing the width of the fracture process zone. 

The strain energy density, $\psi$, is decomposed into elastic and plastic contributions. Since only tensile elastic deformations are presumed to contribute to damage, the elastic strain energy density is further divided into damaging and non-damaging parts.  The final expression of the strain energy density is given by
\begin{equation}
	\psi = g(\phi) \psi^{e+} + \psi^{e-} + g^p(\phi) \psi^p,
	\label{eq:strainenergy}
\end{equation}
where the $+$ and $-$ superscripts denote, respectively, the damaging and non-damaging parts and $g$ denotes the elastic degradation function. Furthermore, this formulation follows the approach by Borden \textit{et al.}\ \cite{Borden2016}, in which the plastic energy density function and yield surface are degraded by the plastic degradation function $g^p$.

A hydrostatic-deviatoric strain split is applied, yielding \cite{Amor2009}
\begin{equation}
	\psi^{e+} = \frac{\kappa}{2} \left< \varepsilon^e_\text{vol} \right>^2 + G \bm{\varepsilon}^e_\text{dev} : \bm{\varepsilon}^e_\text{dev}
	\label{eq:elasenergydam}
\end{equation}
and
\begin{equation}
	\psi^{e-} = \frac{\kappa}{2} \left< -\varepsilon^e_\text{vol} \right>^2.
	\label{eq:elasenergynondam}
\end{equation}

The same power law hardening plasticity model, given by \Cref{eq:plasticflow,eq:yieldsurface,eq:plasticconstraints}, is adopted. Consequently, the plastic strain energy density can be expressed as
\begin{equation}
	\psi^p = \frac{\sigma_{y0}^2}{E(n+1)} \left( 1 + \frac{E (\varepsilon^p_\text{eq}+\varepsilon^{tp}_\text{eq})}{\sigma_{y0}} \right)^{n+1} - \frac{\sigma_{y0}^2}{E(n+1)}.
	\label{eq:plasticenergy}
\end{equation}
Note that the equivalent TRIP strain $\varepsilon^{tp}_\text{eq}$ does not evolve during the fracture simulation stage of the model. Its value is set based on the results from the welding process stage.

Quadratic degradation functions are adopted here to degrade the damaging elastic and plastic parts of the strain energy density, given by
\begin{equation}
	g(\phi) = (1-\phi)^2
	\label{eq:degradation}
\end{equation}
and, following \cite{Mandal2024},
\begin{equation}
	g^p(\phi) = \beta g - \beta + 1.
	\label{eq:degradationplastic}
\end{equation}
Here, $\beta$ represents a parameter that determines the fraction of plastic work contributing to damage. The remaining plastic work is assumed to be dissipated as heat. Based on the work by Taylor and Quinney \cite{Taylor1934}, and following Mandal \textit{et al.}\ \cite{Mandal2024}, $\beta$ is taken as 0.1.

Variational minimization of functional $\Psi$ (\Cref{eq:functional}) and substituting \Cref{eq:crackfunc,eq:strainenergy,eq:elasenergydam,eq:elasenergynondam,eq:plasticenergy,eq:degradation,eq:degradationplastic} yields the following local balance equations
\begin{equation}
	\nabla \cdot \bm{\sigma}  = \mathbf{0},
\end{equation}
and
\begin{equation}
	\frac{G_c}{\ell} \left( \phi - \ell^2 \nabla^2 \phi \right) = -\frac{dg}{d\phi} (\mathcal{H} + \beta \psi^{p} ),
	\label{eq:localbalance_phi}
\end{equation}
with
\begin{equation}
	\bm{\sigma} =  g(\phi) \kappa \left< \varepsilon^e_\text{vol} \right> - \kappa \left< -\varepsilon^e_\text{vol} \right>  + 2g(\phi)G \bm{\varepsilon}^e_\text{dev}.
\end{equation}
Replacing $\psi^{e+}$ in the derivation of \Cref{eq:localbalance_phi}, a history crack driving energy, $\mathcal{H}$, is introduced to enforce damage irreversibility:
\begin{equation}
	\mathcal{H} = \max_{\tau\in[0,t]}(\psi^{e+}).
\end{equation}
Finally, the degraded yield surface related to the degraded plastic strain energy density is given by
\begin{equation}
	f = \sigma_\text{eq} - g^p(\phi) \sigma_{y0}\left(1 + \frac{E (\varepsilon^p_\text{eq}+\varepsilon^{tp}_\text{eq})}{\sigma_{y0}}\right)^n.
\end{equation}

\subsubsection{Hydrogen diffusion}

The transport of hydrogen through the material is described by using the diffusible hydrogen concentration, $C$, as variable. The diffusion process is modeled through Fick's law:
\begin{equation}
	\frac{\partial C}{\partial t} = -\nabla \mathbf{J},
\end{equation}
with the hydrogen flux given by
\begin{equation}
	\mathbf{J} = -D \nabla C + \frac{DC}{RT} V_h \nabla \sigma_h,
\end{equation}
where $D$ is the apparent diffusivity coefficient, $V_h$ is the partial molar volume of hydrogen, and $\sigma_h=\tfrac{1}{3} \text{trace}(\bm{\sigma})$ is the hydrostatic stress. Here, we do not explicitly model hydrogen trapping, the sequestration of hydrogen atoms at microstructural defects \cite{cupertino2023hydrogen}, but its effect is accounted for through the apparent diffusivity constant $D$. Models explicitly accounting for different trapping sites, such as the one described by Isfandbod and Mart\'{\i}nez-Pa\~neda \cite{Isfandbod2021}, can be employed in the future to refine the current simulation framework.

The local hydrogen concentration is assumed to degrade the local critical energy release rate \cite{Martinez-Paneda2018}. Following Mandal \textit{et al.}\ \cite{Mandal2024}, a phenomenological degradation law is used, given by
\begin{equation}
	G_c(C) = f(C) G_{c0} = 
        \left[ f_\text{min} 
	+ \left(1-f_\text{min}\right) 
	\exp(-q_1 C^{q2})
	\right] G_{c0}.
 \label{eq:degradationfunction}
\end{equation}

\subsubsection{Microstructure-based properties}
\label{section:fractureproperties}

In the fracture simulation stage, the material properties are defined based on the phase fraction fields resulting from the welding stage of the framework. Yield strengths for each phase are obtained from the thermodynamic software. Additionally, the bainite phase uses a lower hardening exponent, $n$, and a lower hydrogen diffusivity coefficient, $D_0$, relative to the ferrite and pearlite phases.

Reliable phase-specific parameters for the phase field fracture model are essential for realistic fracture predictions. However, these are not readily available in thermodynamic software and obtaining fracture toughness data for individual constituents experimentally is challenging. Therefore, they are calibrated on crack growth resistance (J-R) curves of pipeline steels. Fracture resistance curves are available, in air and in H$_2$ environments, for base metal samples of pipeline steel grades that contain distinct microstructures (pearlite, ferrite, bainite), allowing us to define $G_c$, and its dependence with hydrogen, as a function of the phase fraction. 

\begin{figure}[!tb]
        \begin{subfigure}{\linewidth}
		\centering
		\includegraphics[width=\linewidth]{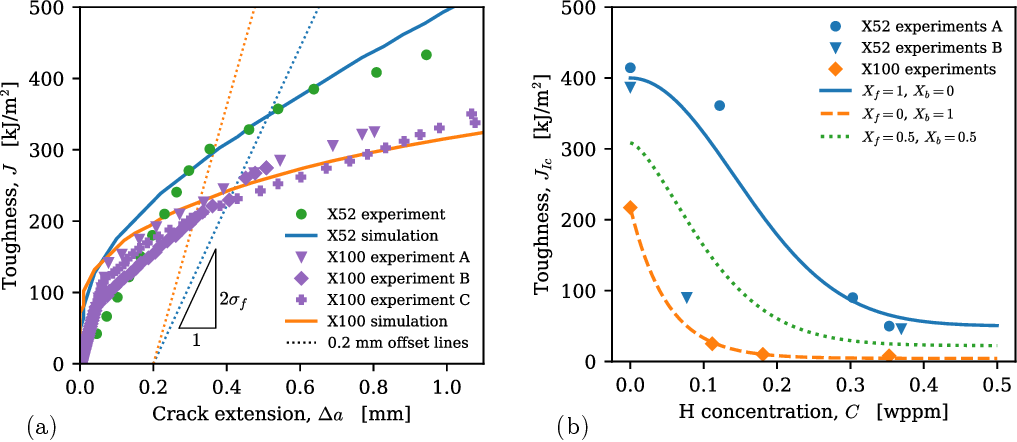}
            \phantomcaption \label{fig:Gdegradation_a}
            \phantomcaption \label{fig:Gdegradation_b}
	\end{subfigure}
	\caption{Fracture toughness data and model fits. (a) J-R curves fitted on experimental data of X52 and X100 grades \cite{Ronevich2018,Ronevich2021} (b) Hydrogen degradation law fitted on experimental data of X52 and X100 grades \cite{Ronevich2021,SanMarchi2021}.}
	\label{fig:Gdegradation}
\end{figure}

The determination of fracture properties of individual phases in an X52 pipeline is demonstrated here, as this material is used in the defect analysis presented in \Cref{section:application}. First, fracture toughness data of the X52 base material, obtained by compact tension (CT) tests by Ronevich \textit{et al.}\ \cite{Ronevich2021} and San Marchi \textit{et al.}\ \cite{SanMarchi2021}, are considered. The critical energy release rate in air, $G_{c0}$, and the length scale parameter, $\ell$, were calibrated on the J-R curve, as shown in \Cref{fig:Gdegradation_a}. The numerical J-R curves were obtained using a boundary layer model in combination with Williams crack tip field solutions \cite{Williams1957}. $G_{c0}$ determines the onset of a numerical J-R curve, while $\ell$ affects the slope, through its relation to the fracture strength \cite{kristensen2020jmps}. \Cref{fig:Gdegradation_b} shows the fitted hydrogen degradation function of \Cref{eq:degradationfunction} against the fracture toughness data. Since the microstructure of the X52 base material consists of a ferrite and pearlite, the fracture parameters calibrated on the X52 data are used for both the ferrite and the pearlite phases. The fracture resistance of the bainite phase is characterised following the experiments by Ronevich \textit{et al.}\ \cite{Ronevich2018,Ronevich2021} on X100 pipeline steel, as the microstructure of this steel primarily consists of bainite. \Cref{fig:Gdegradation} shows the data and model fits for the X100 pipeline steel.

The fitted fracture model parameters for both steels are provided in \Cref{tab:fractureparam}. It should be noted that the initial critical energy release rate of the model, $G_{c0}$, corresponds to the onset of the J-R curves of \Cref{fig:Gdegradation_a}. However, fracture toughness measurements, including the data of \Cref{fig:Gdegradation_b}, commonly report the 0.2 mm crack offset value, $J_{Ic}$, in accordance with standards \cite{ASTM1820}. The 0.2 mm offset lines are depicted as dotted lines in \Cref{fig:Gdegradation_a}, with $J_{Ic}$ defined as the intersection between the J-R curve and this offset line. Therefore, $G_{c0}$ does not directly correspond to the $J_{Ic}$. For example, the fitted $G_{c0}$ value for X100 is higher than that of X52. However, the J-R curve of X100 already falls below that of X52 for minimal crack extension values. This illustrates that the elastoplastic fracture behavior is influenced by multiple material properties, such as yield stress. It is also worth noting that the current fracture model calibration, which uses the properties of an X100 base metal with a bainitic microstructure to represent the bainite formed during the welding process of an X52 pipeline, is somewhat limited. The X100 steel has most likely undergone several postprocessing steps to improve its properties. Therefore, it is expected that the bainite formed in during the welding process has worse fracture properties. This highlights the importance of experimental efforts in characterizing fracture properties of specific HAZ and WM microstructures.

\begin{table}[!tb]
\centering
\caption{Model parameters for the material behavior of X52 and X100 grades of \Cref{fig:Gdegradation}.}
\label{tab:fractureparam}
\begin{tabular}{@{}lccccccc@{}}
    \toprule
    Grade & $\sigma_{y0}$ [MPa] & $n$ [-]  & $G_{c0}$ [kJ/m$^2$] & $\ell$ [mm] & $f_\text{min}$ [-] & $q_1$ [-] & $q_2$ [-] \\ \midrule
    X52   & 416           & 0.1  & 60       & 0.46  & 0.26               & 25   & 2    \\
    X100  & 820           & 0.05 & 80       & 0.35  & 0.10                & 20   & 1    \\ \bottomrule
\end{tabular}
\end{table}

A rule-of-mixtures, similar to \Cref{eq:ruleofmixtures}, is applied to all fracture properties when multiple phases are present. This includes the parameters for the hydrogen degradation function. An example is shown in \Cref{fig:Gdegradation_b}, where the toughness degradation for a microstructure with 50\% ferrite and 50\% bainite is plotted. This approach, which allows for different degradation behaviors based on microstructural composition, provides a valuable means of capturing heterogeneity in the weld region under different hydrogen environments.

\subsubsection{Boundary conditions for pipelines subjected to internal pressure}
\label{section:pressurebcs}

To simulate internal pressure inside the pipeline, a radial displacement, $u^*_r$, is applied to the inner surface of the pipe. This displacement is related to the target pressure, $p$, through the linear elastic relationship for thin-walled cylinders, given by
\begin{equation}
    u^*_r = \frac{pR^2}{bE} \left( 1 - \frac{v}{2} \right),
\end{equation}
where $b$ is the pipeline wall thickness, $R$ is the inner radius, $E$ is Young's modulus, and $\nu$ is Poisson's ratio. It should be noted that the above relation becomes less accurate close to the failure point of the pipeline, where plastic deformations begin to play a role. Therefore, the failure pressure of a pipeline is calculated based on the circumferential (hoop) stress, $\sigma_{\theta\theta}$, at a point in the center of the pipeline wall, far away from the weld, i.e. $p = \sigma_{\theta\theta}b/R$.

When the pipeline is pressurized with hydrogen, Sievert's law is used to prescribe the hydrogen concentration at the inner surface of the pipeline as a function of the target pressure:
\begin{equation}
    C^*_\text{inner} = S\sqrt{p},
\end{equation}
with the hydrogen solubility in steel taken as $S=0.077$ wppm MPa$^{-0.5}$ \cite{Martin2020}. The outer surface of the pipeline, assumed to be free from hydrogen exposure, is prescribed a zero hydrogen concentration, i.e.\ $C^*_\text{outer}=0$.

In the simulations presented in this paper, the target pressure is slowly increased over a timespan of several months. This gives the hydrogen sufficient time to diffuse through the pipeline and reach a steady-state condition. This is a realistic approach for the application at hand, as steady-state conditions are reached in a few days on steel pipelines and these are expected to be in service for decades.

\subsection{Finite element implementation}

\begin{figure}[!tb]
	\centering
	\includegraphics[width=.8\linewidth]{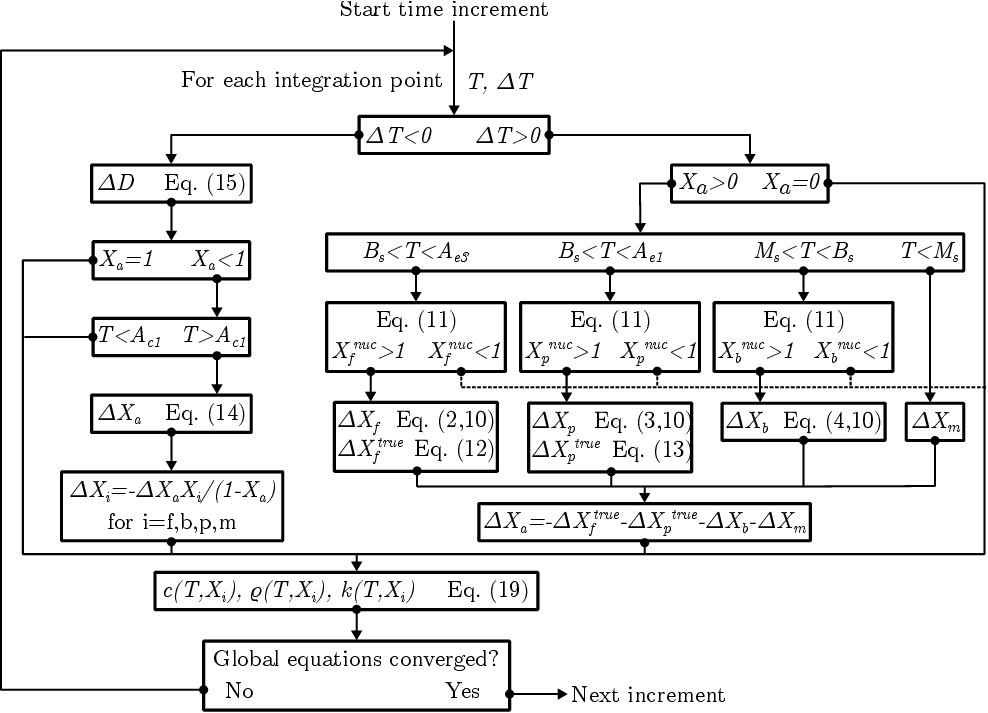}
	\caption{Outline of the implementation of the coupled metallurgical-thermal model at integration point level.}
	\label{fig:thermal_model}
\end{figure}

The simulation framework is implemented in the finite element program ABAQUS. For stage 1, a sequentially-coupled thermal-mechanical analysis is used. The coupled thermal-metallurgical model was implemented via a UMATHT user-subroutine. User-subroutines UEXTERNALDB and USDFLD are used for data initialization and communication. Each integration point is assumed to have an underlying microstructure; the phase transformations of this microstucture are based on the thermal cycle that the integration point experiences. The model implementation at integration point level is depicted in the schematic diagram shown in \Cref{fig:thermal_model}. All rate equations are integrated implicitly using the backward Euler method. Note that there is a two-way coupling between the temperature and phase transformations since phase transformations affect the material properties. The mechanical model, calculating the residual stresses, was implemented in a UMAT user-subroutine, where the temperature and phase fractions from the thermal-metallurgical analysis are transferred via predefined fields. The elastoplastic phase field fracture model coupled to hydrogen diffusion of stage 2 was implemented via a UEL user-subroutine. The developed codes are freely provided to the community and can be downloaded from \href{https://mechmat.web.ox.ac.uk/codes}{https://mechmat.web.ox.ac.uk/codes} \footnote{Code to be uploaded shortly after acceptance of the paper}

In the current study, the framework is applied to weld geometries that can be simplified to 2D. To analyze more complex weld geometries, full 3D pipeline geometries should be used. The implementation of the presented framework in 3D is straightforward but out of the scope of the current study.

\section{Weld process model demonstration and validation}
\label{section:validation}

\subsection{Pipeline geometry and welding cycles}

The outputs of the welding process model are validated against experimental data for seam welds, manufactured with a double submerged arc welding process, on modern and vintage X60 pipeline steels. Here, vintage refers to pipelines that have been in place before the 1970s. Macrographs and microhardness maps are available for both welds. The compositions of the steels are presented in \Cref{tab:x60_composition}. The most significant difference between the two grades is the carbon content, which is notably lower in the modern grade. Consequently, the modern grade also has a significantly lower carbon equivalent (CE), which indicates a reduced tendency to form hard phases upon cooling. 

\begin{table}[!tb]
\centering
\caption{Composition of the X60 vintage and modern steels. Only the alloying elements where wt$\% > 0.1$ are listed. The carbon equivalent (CE) is calculated following the International Institute of Welding (IIW) equation.}
\label{tab:x60_composition}
\begin{tabular}{@{}llllll@{}}
\toprule
Grade       & C [wt\%] & Si [wt\%] & Mn [wt\%] & CE$_\text{IIW}$ \\ \midrule
X60 vintage & 0.176 & 0.217 & 1.37 & 0.42            \\
X60 modern  & 0.071 & 0.233 & 1.25 & 0.29            \\ \bottomrule
\end{tabular}
\end{table}

\begin{figure}[!tb]
	\begin{subfigure}{\linewidth}
		\centering
		\includegraphics[width=\linewidth]{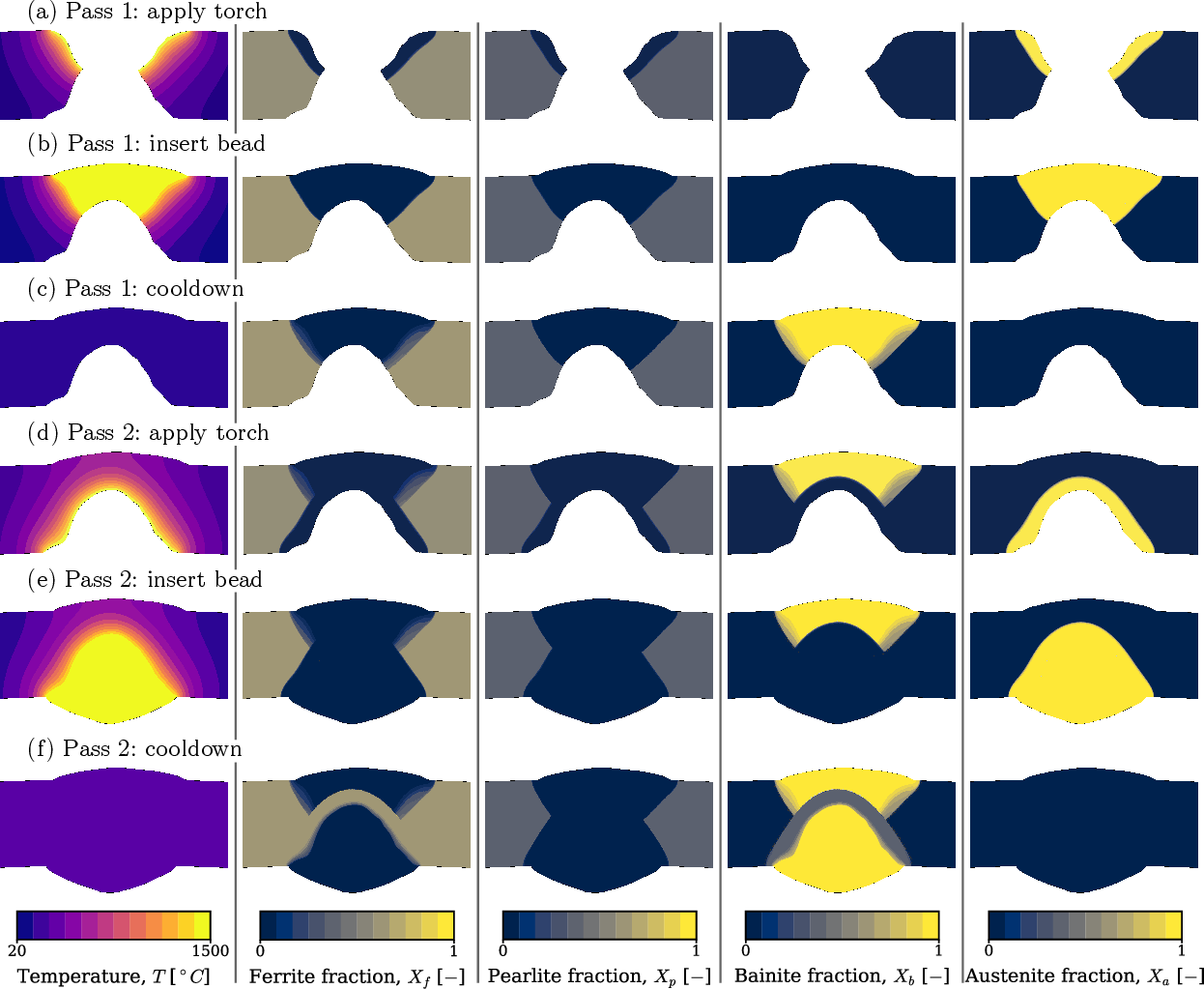}
		\phantomcaption \label{fig:weldcycle_a}
		\phantomcaption \label{fig:weldcycle_b}
		\phantomcaption \label{fig:weldcycle_c}
		\phantomcaption \label{fig:weldcycle_d}
		\phantomcaption \label{fig:weldcycle_e}
		\phantomcaption \label{fig:weldcycle_f}
	\end{subfigure}
	\caption{Temperature profile and phase fraction fields during different stages of a welding process simulation. (a) A torch is applied to the fusion line of the first weld bead, of which the temperature is gradually increased to 1500 $^\circ$C over a period of 3 seconds. (b) The first welding bead is inserted at a temperature of 1500 $^\circ$C.  (c) A cooldown period of 120 seconds reduces the temperature in the weld. (d,e,f) The same process is repeated for the second weld bead.}
	\label{fig:weldcycle}
\end{figure}

The weld geometries in the models are based on the macrographs. A similar approach as described by Mandal \textit{et al.}\ \cite{Mandal2024} and Bensari \textit{et al.}\ \cite{Bensari2019} was used to simulate the welding process. First, the weld beads were removed from the model. Next, the temperature along the fusion lines of the first weld bead was gradually increased to 1500 $^\circ$C over a period of 3 seconds. \Cref{fig:weldcycle_a} displays the temperature distribution and ferrite, pearlite, bainite, and austenite phase fractions at the end of this step for the X60 vintage pipeline weld. It can be seen that the microstructure of the base material, originally consisting of ferrite and pearlite, transforms to austenite near the fusion line. Next, the weld bead was inserted with an initial temperature of 1500 $^\circ$C, as shown in \Cref{fig:weldcycle_b}. The final step of a weld pass is a cooldown period of 120 seconds. \Cref{fig:weldcycle_c} shows that, by the end of this period, a bainitic microstructure was predicted in the WM, while the HAZ microstructure consisted of ferrite and bainite. 

The above steps were repeated for the second weld pass, as shown in \Cref{fig:weldcycle_d,fig:weldcycle_e,fig:weldcycle_f}. The final microstructure in the WM consists of bainite, while the HAZ microstructure has a mixture of ferrite and bainite. In the HAZ adjacent to the first weld bead, the bainite fraction is higher than that of ferrite. The HAZ adjacent to the second weld bead has a slightly lower bainite fraction compared to the HAZ adjacent to the first weld bead. Furthermore, during the second weld pass, a region of the first weld bead was reheated above the austenitization temperature, resulting in a ferrite-bainite microstructure in this region as well. The duration of the torch application periods and intermediate cooldown period were chosen as per the experimental data. The final cooldown period was significantly longer so that the entire pipeline has an uniform temperature close to 20 $^\circ$C.

\subsection{Hardness predictions}

To validate the welding process model predictions, the phase fraction fields were converted into microhardness maps, which can be directly compared to experimentally obtained microhardness maps. 

Maynier \textit{et al.}\ \cite{Maynier1978} developed empirical relations to estimate the hardness of individual phases based on composition and cooling rate. These are given as
\begin{equation}
\begin{split}
	H_{\mathcal{V}f+p} = 42 + 223\text{C} + 53\text{Si} + 30\text{Mn} + 12.6\text{Ni} + 7\text{Cr} + 19\text{Mo} 
	\\+ \left( 10 - 19\text{Si} + 4\text{Ni} + 8\text{Cr} + 130\text{V}\right)\log_{10}[\text{CR}_{700}]
\end{split}
\end{equation}
and
\begin{equation}
\begin{split}
	H_{\mathcal{V}b} = -323 + 185\text{C} + 330\text{Si} + 153\text{Mn} + 65\text{Ni} + 144\text{Cr} + 191\text{Mo} 
	\\ + \left( 89 + 53\text{C} - 55\text{Si} - 22\text{Mn} - 10\text{Ni} + 20\text{Cr} - 33\text{Mo} \right)\log_{10}[\text{CR}_{700}],
\end{split}
\end{equation}
where $\text{CR}_{700}$ represents the cooling rate at 700 $^\circ$C in $^\circ$C/hour. A linear rule of mixtures was used to predict the hardness of a microstructure containing multiple phases:
\begin{equation}
	H_{\mathcal{V}} = (X_f + X_p) H_{\mathcal{V}f+p} + X_b H_{\mathcal{V}b}.
\end{equation}

To demonstrate how the cooling rates affect the hardness, the hardness values for different, but constant, cooling rates are denoted in the CCT diagram of \Cref{fig:x60_cct}.

\begin{figure}[!tb]
	\begin{subfigure}{\linewidth}
		\centering
		\includegraphics[width=.8\linewidth]{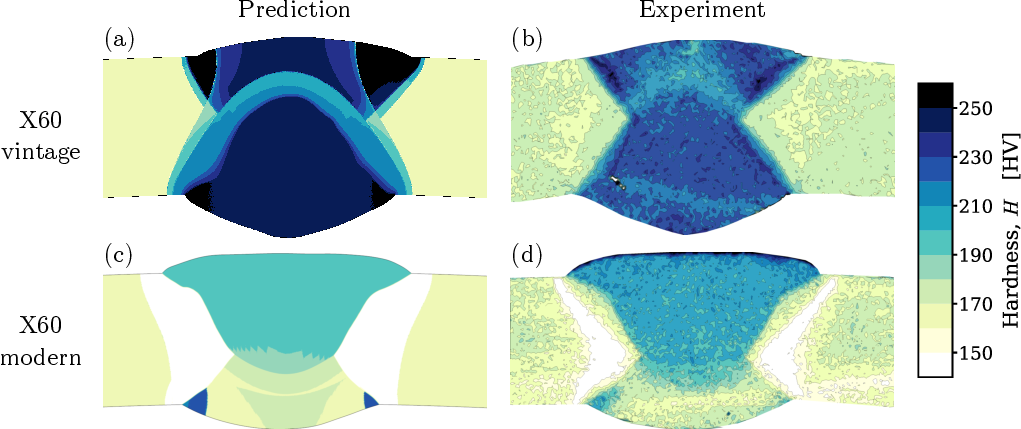}
		\phantomcaption \label{fig:hardness_a}
		\phantomcaption \label{fig:hardness_b}
		\phantomcaption \label{fig:hardness_c}
		\phantomcaption \label{fig:hardness_d}
	\end{subfigure}
	\caption{Comparison of microhardness map predictions with experiments. (a) shows the prediction for an  X60 vintage pipeline steel, while (b) shows the measured microhardness map.  (c) and (d) show, respectively, the predicted and experimental microhardness maps of an X60 modern pipeline steel.}
	\label{fig:hardness}
\end{figure}

The predicted microhardness map for the X60 vintage pipeline steel, derived from the phase fractions of \Cref{fig:weldcycle}, is shown in \Cref{fig:hardness_a}. \Cref{fig:hardness_b} shows the experimentally obtained microhardness map for this weld. The simulation results show a good agreement with experimental data, with predicted hardness values closely matching the measured values. The weld metal is harder than the base material. In the experiment, the center of the top weld bead is slightly softer than the regions close to the fusion line, a feature also captured by the simulation.

The predicted and experimentally obtained microhardness maps for the X60 modern pipeline weld are shown in \Cref{fig:hardness_c,fig:hardness_d}, respectively. For this weld, the bottom bead was created in the first pass and the top bead in the second pass. Unfortunately, the composition of the filler material is not known. A slightly higher carbon content than the base material was assumed, which resulted in the best agreement between prediction and experiment throughout the WM and HAZ. The HAZ zone of this weld is relatively soft. The bottom weld bead exhibits a similar hardness as the base material, while the regions close to the weld toes are harder. The top weld bead is harder than the bottom weld bead.

Overall, the HAZ and WM of the X60 modern grade are softer than the X60 vintage grade. This can be attributed to the lower carbon equivalent of the modern steel. The strong agreement between simulation and experimental results for both welds confirms the validity of the implemented solid-state phase transformation models into the welding process framework.

\subsection{Residual stress predictions}

The accuracy of the residual stress predictions of the framework was validated on a 16-pass girth weld. These results are briefly discussed in \ref{section:stressverification}. This section presents an analysis of the effect of welding input parameters on residual stress profiles to further showcase the capabilities of the framework. For this purpose, the X60 modern pipeline weld, whose microhardness map was discussed in the previous section, is considered.

\begin{figure}[!tb]
	\begin{subfigure}{\linewidth}
		\centering
		\includegraphics[width=\linewidth]{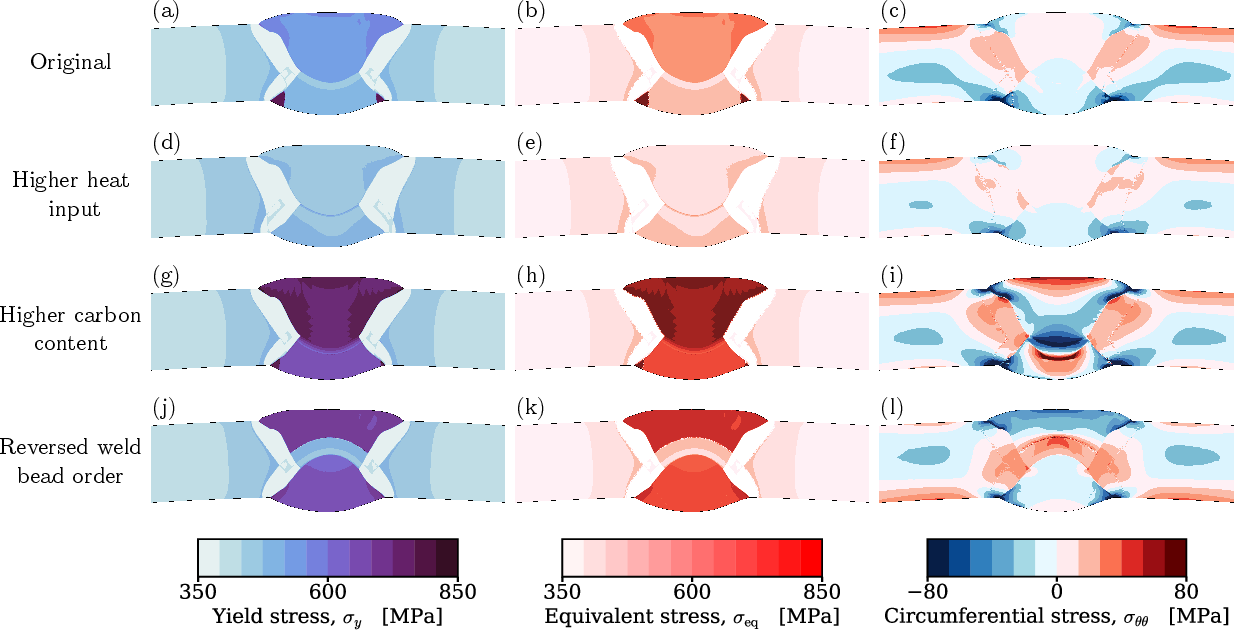}
		\phantomcaption \label{fig:stress_a}
		\phantomcaption \label{fig:stress_b}
		\phantomcaption \label{fig:stress_c}
		\phantomcaption \label{fig:stress_d}
		\phantomcaption \label{fig:stress_e}
		\phantomcaption \label{fig:stress_f}
		\phantomcaption \label{fig:stress_g}
		\phantomcaption \label{fig:stress_h}
            \phantomcaption \label{fig:stress_i}
            \phantomcaption \label{fig:stress_j}
            \phantomcaption \label{fig:stress_k}
            \phantomcaption \label{fig:stress_l}
	\end{subfigure}
	\caption{The effect of different welding strategies on residual stresses. (a,d,g,j) Yield stress, (b,e,h,k) equivalent von Mises stress, and (c,f,i,l) residual circumferential stress for the (a,b,c) original simulation, (d,e,f) higher heat input, (g,h,i) weld filler material with higher carbon content, and (j,k,l) reversed order of the weld beads.}
	\label{fig:stress}
\end{figure}

The yield stress (including hardening), equivalent von Mises stress, and residual circumferential stress at the end of the original simulation are respectively presented at \Cref{fig:stress_a,fig:stress_b,fig:stress_c}. The equivalent stress distribution largely follows the yield stress profile, with values close to the yield stress throughout the entire weld. High residual tensile stresses are observed in the HAZ. The regions near the weld toes experience relatively high compressive stresses. In the center of the WM, slightly lower tensile and compressive stresses are observed.

To demonstrate the possibilities of the welding process model, three additional simulations of the same weld, but under different welding conditions, were conducted. \Cref{fig:stress_d,fig:stress_e,fig:stress_f} show results of a simulation with a higher heat input, where the weld beads were inserted at a temperature of 1800 $^\circ$C. In this scenario, the yield stress of the weld metal is reduced compared to the original case. This reduction is attributed to a slightly lower bainite fraction resulting from a slower cooling rate. Due to the increased heat input, more heat has to dissipate from the weld, leading to this decrease in cooling rate. The residual stress in the HAZ is lower than in the original simulation, while the high compressive stress regions around the bottom weld toes have become smaller. 

The results presented in \Cref{fig:stress_g,fig:stress_h,fig:stress_i} were obtained using a filler material with a carbon content of 0.25 wt\%, compared to 0.15 wt\% in the original simulation. The higher carbon results in increased hardenability of the material, which is reflected by the high yield stress in the WM. Consequently, the tensile stresses in the HAZ have increased. Furthermore, the center of the WM now exhibits alternating high tensile and compressive stresses.

In the final case, shown in \Cref{fig:stress_j,fig:stress_k,fig:stress_l}, the order in which the weld beads are laid is reversed compared to the original simulation. The resulting yield stress in the WM is higher. This can be attributed to the difference in size between the two weld beads. The larger top weld bead introduced more heat when inserted. When this weld bead is inserted last (as in the original case), the cooling rate during the final pass is lower than when the smaller weld bead is laid last (as in the reversed case). As a result, the WM in the original case is softer than in the reversed case. The residual circumferential stress prediction in \Cref{fig:stress_h} shows a high tensile region in the center of the WM, which is not present in the other simulations. 

The analysis presented in this section demonstrates that choices made during the welding process can influence the hardness and residual stresses within the welds and HAZ. Furthermore, it showcases the capabilities of the computational framework to explore different welding strategies.

\section{Failure predictions of an X52 pipeline}
\label{section:application}

\begin{figure}[!tb]
	\begin{subfigure}{\linewidth}
		\centering
		\includegraphics[width=\linewidth]{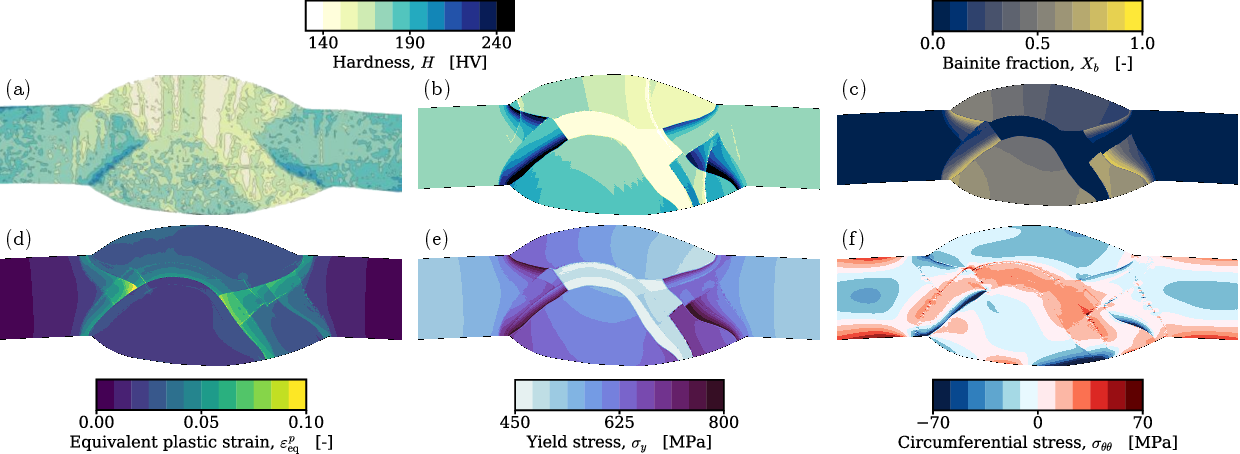}
		\phantomcaption \label{fig:b50weld_a}
		\phantomcaption \label{fig:b50weld_b}
		\phantomcaption \label{fig:b50weld_c}
		\phantomcaption \label{fig:b50weld_d}
		\phantomcaption \label{fig:b50weld_e}
		\phantomcaption \label{fig:b50weld_f}
	\end{subfigure}
	\caption{Results of the welding process model during stage 1 of the framework. (a) Experimental microhardness map. (b-f) Results obtained from the welding process stage of the simulation framework, with (b) the microhardness map, (c) the bainite fraction, (d) the equivalent plastic strain, (e) the yield stress, and (f) the residual circumferential stress.}
	\label{fig:b50weld}
\end{figure}

In this section, we analyze the effect of defects located at various positions on the failure pressure of the pipeline. Specifically, we investigate a seam weld of an X52 pipeline manufactured in 1950 through a double submerged arc welding process, which was recently taken out of service. The pipeline has an outer diameter of 762 mm with a wall thickness of 8.2 mm. An experimental analysis of this weld was presented in Ref. \cite{Agnani2023}. The experimentally obtained microhardness map, presented in \Cref{fig:b50weld_a}, reveals hard regions within the HAZs adjacent to the bottom weld beads. Such hard regions are commonly considered problematic for the structural integrity of the weld. It should be noted that this weld was manufactured using three passes; a traditional submerged arc weld that was likely repaired during original pipe fabrication with an additional weld pass.

After a mesh sensitivity study, the pipeline was discretized by 156,512 quadratic quadrilateral elements, resulting in 174,266 nodes. An element size of 0.06 mm was adopted in the weld so that there are at least five elements within the phase field length scale \citep{kristensen2021assessment}. Far away from the weld, a much coarser element size was used. Furthermore, a plane strain condition was assumed in the simulations. 

\subsection{Weld process}

The results from the welding process stage of the modeling framework are displayed in \Cref{fig:b50weld_b,fig:b50weld_c,fig:b50weld_d,fig:b50weld_e,fig:b50weld_f}. The predicted microhardness map of \Cref{fig:b50weld_b} shows reasonable agreement with the experimental map. The hard HAZ regions, which are our focus here, are slightly more pronounced in the predictions. \Cref{fig:b50weld_c} reveals high bainite fractions in the hard HAZs, with the WM microstructure also containing a considerable amount of bainite. Most plastic deformation, as shown in \Cref{fig:b50weld_d}, occurs in the HAZs, leading to strain hardening in these regions. This strain hardening is reflected in the yield stress distribution in \Cref{fig:b50weld_e}. The residual circumferential stress distribution in \Cref{fig:b50weld_f} reveals high tensile stress regions near the left-bottom HAZ and the center of the WM. This high tensile stress is not present in the HAZ located on the right side of the weld.

\subsection{Fracture toughness distribution}

\begin{figure}[!tb]
	\begin{subfigure}{\linewidth}
		\centering
		\includegraphics[width=\linewidth]{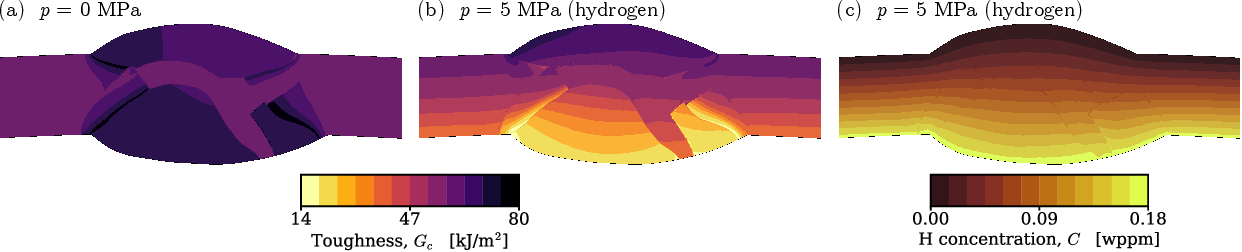}
		\phantomcaption \label{fig:Gfield_a}
		\phantomcaption \label{fig:Gfield_b}
		\phantomcaption \label{fig:Gfield_c}
	\end{subfigure}
	\caption{Toughness distribution in the simulation for (a) zero hydrogen pressure (or any air pressure) and (b) a hydrogen pressure of 5 MPa. (c) The hydrogen concentration around the weld for a hydrogen pressure of 5 MPa.}
	\label{fig:Gfield}
\end{figure}

The fracture toughness distribution, including degradation effects due to hydrogen, is obtained based on phase fraction predictions from the weld process stage, as outlined in \Cref{section:fractureproperties}. The fracture toughness distribution in air is shown in \Cref{fig:Gfield_a}, displaying relatively uniform toughness across the weld and pipeline. This is attributed to the similar initial critical energy release rates of the X52 ferrite-pearlite microstructure and the X100 bainitic microstructure on which the bainite phase fracture properties are based (see \Cref{section:fractureproperties}). However, it is worth noting that this figure shows the $G_0$ distribution, which represents the onset of the J-R curves in \Cref{fig:Gdegradation_a}, and not the 0.2mm offset fracture toughness $J_{Ic}$, which would be lower in bainite compared to ferrite and pearlite.

Next, the pipeline was pressurized with hydrogen to 5 MPa. The boundary conditions were prescribed as outlined in \Cref{section:pressurebcs}. For illustration purposes, damage was not considered in this simulation. \Cref{fig:Gfield_c} shows the hydrogen concentration distribution in the weld. An approximately linear gradient from the applied hydrogen concentration at the inner surface to a zero hydrogen concentration at the outer surface can be observed. This shows that variations in the diffusivity coefficient, based on the phase fractions, do not play a significant role. Due to the slow loading rate, the hydrogen concentration of \Cref{fig:Gfield_c} reached a steady-state condition. It is expected that diffusivity variations can play a more significant role for faster loading rates. The fracture toughness distribution associated with this hydrogen concentration field is displayed in \Cref{fig:Gfield_b}. It can be observed that the fracture toughness in the hard regions of the HAZ is significantly reduced, which can be attributed to the more severe toughness degradation in the bainite phase.

\subsection{Failure of the defect-free pipeline}

\begin{figure}[!tb]
    \centering
    \includegraphics[width=.5\linewidth]{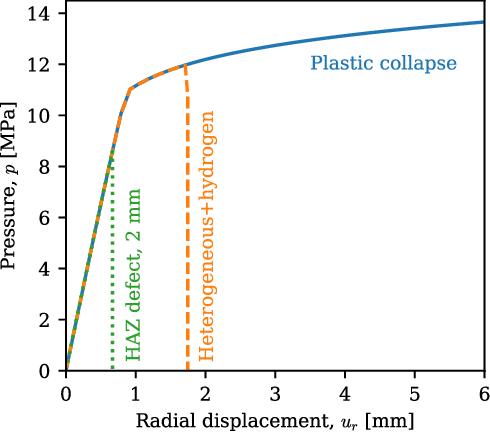}
    \caption{Internal pressure versus radial displacement for various simulations. The plastic collapse curve corresponds to the defect-free simulations with homogeneous properties in both air and hydrogen, shown in \Cref{fig:nodefect_a}, and the defect-free simulation with heterogeneous properties in air, displayed in \Cref{fig:nodefect_b}. The curve labelled as ``Heterogeneous + hydrogen" corresponds to the simulation shown in \Cref{fig:nodefect_c,fig:nodefect_d}. The curve labelled as ``HAZ defect, 2 mm" corresponds to the simulation of \Cref{fig:defects_a} that will be discussed in \Cref{section:defects}.}
    \label{fig:pressuredisp}
\end{figure}

In this section, we examine the effect of heterogeneity on the structural integrity of an initially defect-free pipeline. First, a pipeline with homogeneous properties, equal to those of the base material, is considered. The internal pressure was gradually increased using either a hydrogen-free gas, such as air, or using hydrogen until the maximum attainable pressure was reached. Failure of the pipeline occurred due to plastic collapse. The curve labelled ``Plastic collapse" in \Cref{fig:pressuredisp} shows the pressure as a function of the radial displacement of the inner surface of the pipe. The pressure saturates due to yielding, ultimately reaching a value of approximately 15 MPa.

\begin{figure}[!tb]
	\begin{subfigure}{\linewidth}
		\centering
		\includegraphics[width=\linewidth]{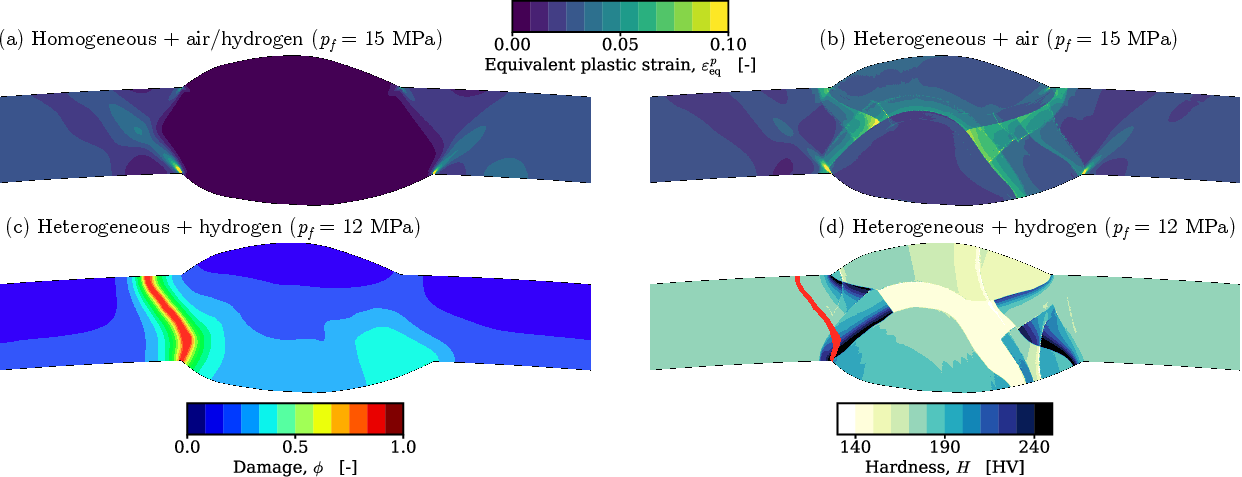}
		\phantomcaption \label{fig:nodefect_a}
		\phantomcaption \label{fig:nodefect_b}
		\phantomcaption \label{fig:nodefect_c}
            \phantomcaption \label{fig:nodefect_d}
	\end{subfigure}
	\caption{Comparison between failure modes for a defect-free pipeline in air and hydrogen. Equivalent plastic strain at the point of plastic collapse for an air-pressurized pipe with (a) homogeneous properties and (b) heterogeneous properties. (c) Damage variable after failure of a pipeline pressurized with hydrogen. (d) Predicted microhardness map with an overlay of the crack path of (c).}
	\label{fig:nodefect}
\end{figure}

The equivalent plastic strain distribution is shown in \Cref{fig:nodefect_a}, where strain peaks at the bottom weld toes can be seen. Additionally, plastic deformation extended throughout the whole pipeline, as evidenced by the non-zero plastic strain values far from the weld. No differences were observed between the air and hydrogen cases; when variations in the underlying microstructure are neglected, the hydrogen degradation is not enough to trigger a shift to brittle fracture. 

Next, heterogeneous material properties within the weld were introduced by using the results from the weld process stage (\Cref{fig:b50weld}) as initial conditions. For the pipeline pressurized with air, the failure pressure (15 MPa) and mode were the same as for the homogeneous case. \Cref{fig:nodefect_b}
 shows the equivalent plastic strain field. Compared to the initial plastic strain field of \Cref{fig:b50weld_d}, additional strain peaks at the bottom weld toes and deformation far from the weld can be seen.

Under hydrogen loading, the pipeline failed due to fracture at a pressure of 12 MPa. The pressure-displacement curve is presented in \Cref{fig:pressuredisp}, labeled as ``Heterogeneous+hydrogen". The curve shows that failure occurred after a small amount of yielding in the pipeline. The phase field variable is shown in \Cref{fig:nodefect_c}. Additionally, the crack profile was extracted from this field and was overlaid onto the microhardness map, as shown in \Cref{fig:nodefect_d}. Severely localized plastic strain at the left-bottom toe resulted in damage initiation and the formation of a small crack. After propagation over a small distance through the HAZ, the crack shifted direction and propagated under an angle of approximately 45$^\circ$ to the outer free surface. This crack propagation was unstable. The heterogeneity within the weld region thus brings a change in failure mode, from plastic collapse to fracture, and reduces the failure pressure when the pipeline is subjected to internal hydrogen pressure, even when the weld region is defect free.

\subsection{On the role of pre-existing defects}
\label{section:defects}

\begin{figure}[!tb]
	\begin{subfigure}{\linewidth}
		\centering
		\includegraphics[width=\linewidth]{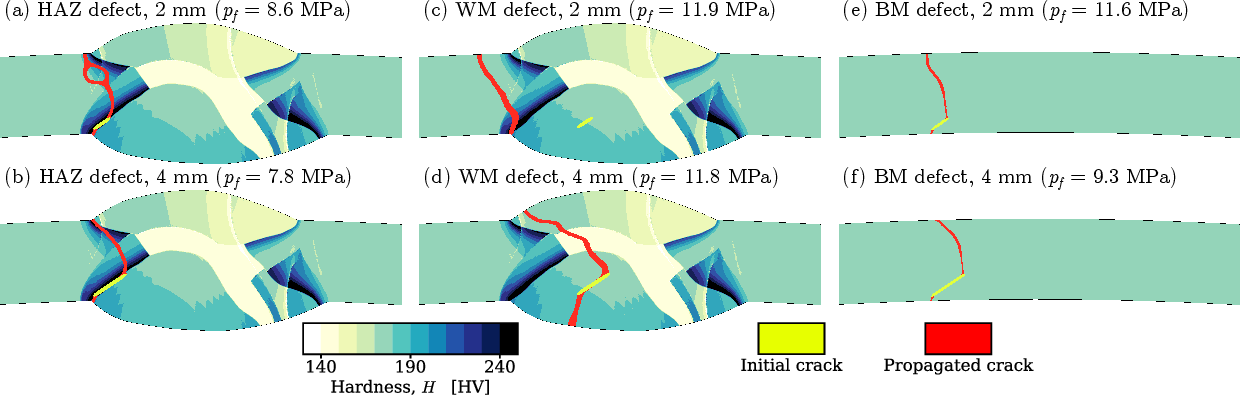}
		\phantomcaption \label{fig:defects_a}
		\phantomcaption \label{fig:defects_b}
		\phantomcaption \label{fig:defects_c}
		\phantomcaption \label{fig:defects_d}
		\phantomcaption \label{fig:defects_e}
		\phantomcaption \label{fig:defects_f}
	\end{subfigure}
	\caption{Comparison of crack paths and failure pressures for different defect locations. Predicted microhardness maps with overlays of initial and propagated cracks located at (a,b) the HAZ, (c,d) the WM, (e,f) the base material, for initial crack lengths of (a,c,e) 2 mm and (b,d,f) 4 mm.}
	\label{fig:defects}
\end{figure}

Initial defects were introduced at three different locations: heat-affected zone, weld metal and base metal. For each defect location, two defect lengths were investigated, namely, 2 mm and 4 mm. Due to the two-dimensional plane strain assumption that is used, the defects are considered to be infinitely long in the out-of-plane direction. This provides a conservative estimation of the failure pressures. The initial defects and their subsequent propagation paths are shown as yellow and red overlays, respectively, in the microhardness maps in \Cref{fig:defects}. 

For the first defect, located in the left-bottom hard HAZ along the weld fusion line (\Cref{fig:defects_a,fig:defects_b}), the hydrogen pressures at failure were significantly reduced to 8.6 MPa and 7.8 MPa for defect lengths of 2 mm and 4 mm, respectively. The pressure-displacement curve for the 2 mm defect is presented in \Cref{fig:pressuredisp}, labeled as ``HAZ defect, 2 mm". The introduced defect has shifted the failure point to the region where the global response of the pipeline is still elastic, resulting in a brittle failure mode. The failure pressure for the 4 mm defect is not much lower than that for the 2 mm defect. This demonstrates that even a small crack in a hard HAZ can pose a significant risk for hydrogen transmission pipelines. 

The propagation of both cracks occurred in two stages. Initially, the crack extended to the inner surface, effectively creating a larger edge crack. However, the pipeline did not yet fail. At a higher pressure, the upper part of the crack propagated through the HAZ towards the left-top weld toe, leading to catastrophic failure.

The second crack was introduced in the bottom weld bead. The crack paths are shown in \Cref{fig:defects_c,fig:defects_d}. The failure pressures of 11.9 and 11.8 MPa for crack lengths of 2 and 4 mm, respectively, were very close to that of the defect-free pipeline of \Cref{fig:nodefect_b}. Notably, failure for the pipeline with the 2 mm defect did not initiate from the defect itself. Instead, the same crack path as for the defect-free case was observed. The 4 mm defect did propagate, although at nearly the same failure pressure. These results suggest that for this particular weld configuration, a defect within the WM poses a lesser risk than the one within the hard HAZ. 

Finally, a defect was introduced in the base material, far from the weld. The defect is located at the same radial distance as the previous two defects. The properties in this area are homogeneous with no residual stress. The crack paths for the 2 mm and 4 mm cracks are shown in \Cref{fig:defects_e,fig:defects_f}, respectively. The failure pressure for the 2 mm defect was 11.6 MPa, which is close to the failure pressure of the defect-free pipeline. For the 4 mm defect, failure occurred at a hydrogen pressure of 9.3 MPa. Interestingly, the failure pressures for this defect location are lower than that for the defect location in the WM of \Cref{fig:defects_c,fig:defects_d}. This can likely be explained by the increased thickness of the weld region, which contributes to the defect tolerance of this region.

The above findings demonstrate the detrimental effect of hard HAZ regions on the defect tolerance of pipelines used for hydrogen transport.

\section{Summary and conclusions}
\label{Sec:Conclusions}

A novel computational framework integrating thermo-metallurgical-mechanical process modeling with phase field-based coupled fracture predictions has been presented. The framework is particularised to the quantification of critical failure pressures in hydrogen transport pipelines, tackling a pressing need, but can be readily applied to other processes and problems, such as additive manufacturing. The model enables incorporating phase transformations into welding process simulations and evaluating their influence on fracture behavior. An excellent agreement is obtained with experimentally-determined microhardness maps, demonstrating the predictive capabilities of the computational framework. As such, the presented framework: (i) enables the prediction of heterogeneous material properties and weak spots within weld regions, (ii) can be used to evaluate the effect of different steel compositions and welding parameters, (iii) eliminates the need for experimental microhardness maps in structural integrity assessment simulations, making the framework applicable to a broad range of situations; and (iv) offers mechanistic insights into the complex relationships between microstructure, residual stresses, and fracture.

The study of the feasibility of repurposing existing pipeline infrastructure for hydrogen transport through the present framework brought important insight. Key findings include:
\begin{itemize}
    \item Heterogeneous material properties within the weld region can lower the failure pressure of hydrogen transportation pipelines and bring a shift in the failure mechanism, from ductile plastic collapse, to catastrophic fracture.  
    \item Hard spots in the heat-affected zone (HAZ), with an underlying bainitic microstructure, are detrimental for the defect tolerance of a pipelines subjected to internal hydrogen pressure.
    \item For realistic conditions and welds, model predictions show that critical failure pressures can be as low as 8.6 MPa in the present of a 2 mm HAZ defect.
\end{itemize}

These findings highlight the critical role of microstructural heterogeneity within weld regions on the structural integrity of hydrogen pipelines.

\section*{Acknowledgments}
\label{Acknowledge of funding}

\noindent The authors acknowledge financial support from EPRI through the R\&D project "Virtual Testing of hydrogen-sensitive components". E.\ Mart\'{\i}nez-Pa\~neda acknowledges financial support from UKRI's Future Leaders Fellowship programme [grant MR/V024124/1]. J.\ Wijnen gratefully acknowledges T.K.\ Mandal (IIT Bombay) for providing data and simulation codes, as well as for stimulating discussions on various topics related to this paper. 

\section*{Data availability statement}

The ABAQUS subroutines used in this work will be made available at \url{https://mechmat.web.ox.ac.uk/codes}. 

\appendix

\section{Residual stress verification}
\label{section:stressverification}
    
\begin{figure}[!tb]
	\begin{subfigure}{\linewidth}
		\centering
		\includegraphics[width=.9\linewidth]{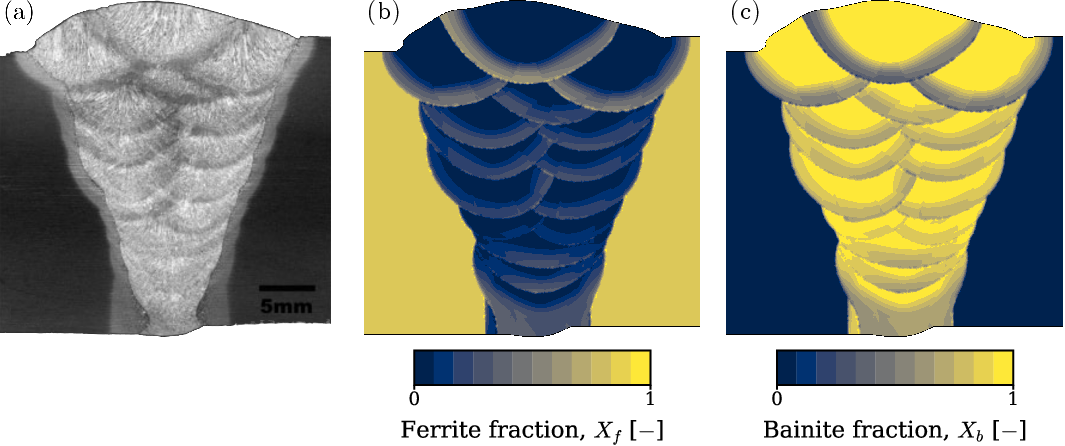}
		\phantomcaption \label{fig:appendix1_a}
		\phantomcaption \label{fig:appendix1_b}
		\phantomcaption \label{fig:appendix1_c}
	\end{subfigure}
	\caption{Geometry of the 16 pass girth weld: (a) macrograph \cite{Neeraj2011}, (b) ferrite phase fraction, (c) bainite phase fraction.}
	\label{fig:appendix1}
\end{figure}

The capability of the framework to accurately predict residual stresses was verified against experimental results presented by Neeraj \textit{et al.}\ \cite{Neeraj2011}, who performed neutron diffraction analyses on girth welds. The considered pipeline is classified as X65, and has an outside diameter of 508 mm and a wall thickness of 25.4 mm. The weld was made with a carbon steel and 16 weld passes. A macrograph of the weld is shown in \Cref{fig:appendix1_a}.

An axisymmetric simulation of the welding process was conducted. The predicted ferrite and bainite fractions are shown in \Cref{fig:appendix1_b,fig:appendix1_c}, respectively. The weld metal microstructure mainly consists of bainite. In each weld pass, small regions of already deposited weld metal were reheated. These regions exhibit a ferritic-bainitic microstructure.

\begin{figure}[!tb]
	\begin{subfigure}{\linewidth}
		\centering
		\includegraphics[width=\linewidth]{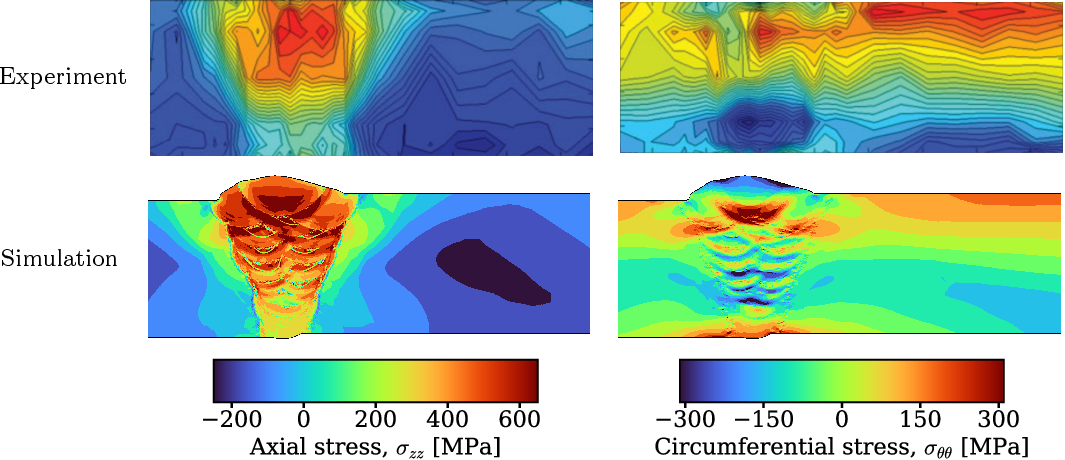}
	\end{subfigure}
	\caption{Measured \cite{Neeraj2011} and predicted residual stress profiles.}
	\label{fig:appendix2}
\end{figure}

The measured and predicted axial and circumferential stress profiles are presented in \Cref{fig:appendix2}. The model accurately captured both the trends and peak values. It should be noted that the experimental stress profiles were obtained on a relatively coarse grid. Consequently, the predicted stress profiles show a much higher level of detail.

\end{document}